\documentclass[twocolumn]{aastex631}

\newcommand{\orcid}[1]{\href{https://orcid.org/#1}
{\includegraphics[width=8pt]{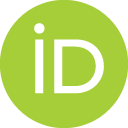}}}
\newcommand{\methane}{CH$_4$}
\newcommand{\cotwo}{CO$_2$}

\usepackage{amsmath}

\received{July 26, 2023}
\revised{September 18, 2023}
\accepted{September 26, 2023}
\published{in ApJ Letters, October 20, 2023}

\begin{document}

   \title{Probing reflection from aerosols with the near-infrared dayside spectrum of WASP-80b}

\correspondingauthor{Bob Jacobs}
\email{b.jacobs@uva.nl}

\author[0000-0002-0373-1517]{Bob Jacobs}
\affiliation{Anton Pannekoek Institute for Astronomy, University of Amsterdam,
Science Park 904, 1098 XH,
Amsterdam, the Netherlands}
\author[0000-0002-0875-8401]{Jean-Michel D\'esert}
\affiliation{Anton Pannekoek Institute for Astronomy, University of Amsterdam,
Science Park 904, 1098 XH,
Amsterdam, the Netherlands}
\author[0000-0002-8518-9601]{Peter Gao}
\affiliation{Earth and Planets Laboratory, Carnegie Institution for Science, 5241 Broad Branch Road, NW, Washington, DC 20015, USA}
\author[0000-0002-4404-0456]{Caroline V. Morley}
\affiliation{The University of Texas at Austin,
Department of Astronomy,
2515 Speedway, Stop C1400,
Austin, Texas 78712-1205}
\author{Jacob Arcangeli}
\affiliation{Anton Pannekoek Institute for Astronomy, University of Amsterdam,
Science Park 904, 1098 XH,
Amsterdam, the Netherlands}
\author[0009-0000-6113-0157]{Saugata Barat}
\affiliation{Anton Pannekoek Institute for Astronomy, University of Amsterdam,
Science Park 904, 1098 XH,
Amsterdam, the Netherlands}

\author{Mark S. Marley}
\affiliation{Lunar and Planetary Laboratory, The University of Arizona, 1629 E. University Blvd.
P.O. Box 210092
Tucson, AZ 85721-0092}
\author{Julianne I. Moses}
\affiliation{Space Science Institute
4765 Walnut St, Suite B
Boulder, CO 80301}
\author{Jonathan J. Fortney}
\affiliation{Department of Astronomy and Astrophysics, 
University of California, 
Santa Cruz, 95064, CA, USA}
\author[0000-0003-4733-6532]{Jacob L. Bean}
\affiliation{Department of Astronomy \& Astrophysics,
University of Chicago,
5640 S. Ellis Avenue,
Chicago, IL 60637,
USA}
\author{Kevin B. Stevenson}
\affiliation{Johns Hopkins APL, 11100 Johns Hopkins Rd, Laurel, MD 20723, USA}
\author{Vatsal Panwar}
\affiliation{Anton Pannekoek Institute for Astronomy, University of Amsterdam,
Science Park 904, 1098 XH,
Amsterdam, the Netherlands}
\affiliation{Department of Physics, University of Warwick, Coventry, CV4 7AL, UK}



\begin{abstract}

The presence of aerosols is intimately linked to the global energy budget and the composition of a planet's atmospheres. Their ability to reflect incoming light prevents energy from being deposited into the atmosphere, and they shape spectra of exoplanets. We observed five near-infrared secondary eclipses of WASP-80b
with the Wide Field Camera 3 (WFC3) aboard the \textit{Hubble Space Telescope} to provide constraints on the presence and properties of atmospheric aerosols.
We detect a broadband eclipse depth of $34\pm10$\,ppm for WASP-80b. We detect a higher planetary flux than expected from thermal emission alone at $1.6\sigma$, which hints toward the presence of reflecting aerosols on this planet's dayside, indicating a geometric albedo of $A_g<0.33$ at 3$\sigma$.
We paired the WFC3 data with Spitzer data and explored multiple atmospheric models with and without aerosols to interpret this spectrum. 
Albeit consistent with a clear dayside atmosphere, we found a slight preference for near-solar metallicities and for dayside clouds over hazes. We exclude soot haze formation rates higher than $10^{-10.7}$\,g\,cm$^{-2}$s$^{-1}$ and tholin formation rates higher than $10^{-12.0}$\,g\,cm$^{-2}$s$^{-1}$ at $3\sigma$. 
 We applied the same atmospheric models to a previously published WFC3/Spitzer transmission spectrum for this planet and found weak haze formation. 
 A single soot haze formation rate best fits both the dayside and the transmission spectra simultaneously. However, we emphasize that no models provide satisfactory fits in terms of the chi-square of both spectra simultaneously, indicating longitudinal dissimilarity in the atmosphere's aerosol composition.


\end{abstract}

   \keywords{planets and satellites: atmospheres --- 
planets and satellites: gaseous planets}


\section{Introduction}

Aerosols are thought to be ubiquitous in exoplanet atmospheres \citep[see, e.g.][]{Sing2011,Kreidberg2014a,Knutson2014,Rustamkulov2023}, although their exact nature remains to be determined. Here, we follow the definitions of \citet{Gao2021} and consider the two main types of aerosols in giant exoplanet atmospheres: clouds and hazes. Haze modeling efforts typically focus on complex hydrocarbon ``soot'' hazes \citep{Lavvas2017} and the Titan analog ``tholin'' hazes \citep{Khare1984}. Clouds form when gas condenses under thermochemical equilibrium, while 
hazes are produced photochemically under strong UV irradiation \citep{ackerman2001, Morley2012, He2018a, Kawashima2018}. 

Aerosol absorption and reflectivity are wavelength-dependent \citep{Adams2019, Feinstein2023}. They therefore modify the thermal structure of the atmosphere \citep{McKay1991, McKay1999, Heng2012, Keating2017}, which, in turn, affects the formation rate of aerosols \citep{Morley2015, Gao2018}. Atmospheric properties such as metallicity, vertical mixing strength, and longitudinal transport also determine the abundance and composition of aerosols \citep{Parmentier2013, Gao2018, He2018a}.

Aerosols attenuate molecular features in near-infrared (NIR) transmission spectra of exoplanets \citep{Fortney2005}. 
Secondary eclipse spectra are likewise shaped by 
aerosols \citep{Demory2013}. Besides probing thermal emission, secondary eclipse depth measurements can also shine a light on the geometric albedo of a planet \citep{Sudarsky2000,Demory2011,Evans2013,Brandeker2022}. Clouds generally create flat NIR reflection spectra, and soot hazes have a low single-scattering albedo. However, tholin particles are highly reflective at optical and NIR wavelengths with strong spectral features while being highly absorbing at blue and UV wavelengths \citep{Morley2015}. Reflectance spectra can therefore help detect aerosols and distinguish their composition.

WASP-80b \citep[$M_p=0.54\,M_{\rm{Jup}}$, $R_p=0.95\,R_{\rm{Jup}}$;][]{Triaud2013,Triaud2015} is a warm Jupiter ($T_{\rm{eq}}\approx800$\,K) orbiting a late-K/early-M dwarf. Its size, proximity to its relatively low-mass host star, and low equilibrium temperature uniquely allow us to probe its NIR reflectance. In addition, because of its host star's high chromospheric activity \citep{Fossati2022}, WASP-80b is expected to receive a high UV flux, which potentially enhances haze production. Therefore, WASP-80b is currently amongst the best targets to investigate NIR reflectance caused by aerosols.

From a theoretical standpoint, the dayside of WASP-80b is in a peculiar thermal parameter space in which its atmosphere could be clear, hazy, or hosting high-pressure silicate clouds \citep{Morley2015,Gao2020}. Its equilibrium temperature is also just below the threshold where the dominating carbon bearer changes from \methane-dominated to CO/\cotwo-dominated \citep{Moses2013,Fortney2020,Baxter2021}. As tholin particles become unstable above $\sim$900\,K \citep{Morisson2016}, hotter hazes are expected to be dominated by more-refractory soot particles \citep{Lavvas2021}. As such, the WASP-80b dayside is also expected to lie in the transitory region between tholin and soot hazes.

\citet{Wong2022} used the \textit{Hubble Space Telescope} (HST) and the Spitzer Space Telescope to measure the WASP-80b transmission spectrum in the range of $0.4-5.0$\,$\mu$m. They observed a muted water feature at $1.4$\,$\mu$m and a steep optical spectral slope. They attribute these features to fine-particle hazes (${<}0.1$\,$\mu$m) and a deep cloud deck on the limbs of the planet. Their models show a slight preference toward tholins over soots at metallicities ranging from ${\sim}30$ to $100$ times solar. General Circulation Models of planets of a similar temperature to WASP-80b predict efficient heat transport and a low day-to-night contrast resulting in chemically homogeneous planets \citep{Showman2015}. The dayside of WASP-80b should therefore also show signs of aerosols.

In this study we present the NIR secondary eclipse spectrum of WASP-80b observed by the HST \textit{Wide Field Camera 3} (WFC3), probing the planet's dayside. We use atmospheric models including aerosol feedback to interpret the spectrum. We also compare the same models to the planet's transmission spectrum \citep{Wong2022}, which probes the planet's limbs.

\section{Data analysis}

\subsection{Observations}
\label{sec:obs}
We observed five secondary eclipses of WASP-80b with four HST orbits per eclipse with WFC3 for Program GO 15131 (PI: J.-M. Désert, DOI:\dataset[10.17909/zr1e-9f27]{https://doi.org/10.17909/zr1e-9f27}). The observation of one eclipse is called a ``visit.'' The data were obtained with the G141 grism, covering $1.1-1.7$\,$\mu$m, using the bidirectional spatial scanning technique. We used the $256\times256$ subarray and the \texttt{SPARS10}, \texttt{NSAMP=14} readout mode. 
An observation log is provided in Table \ref{tab:obs} in Appendix \ref{app:fitting_methods_exp}.

\begin{figure*}
    \centering
    \includegraphics[width=\textwidth]{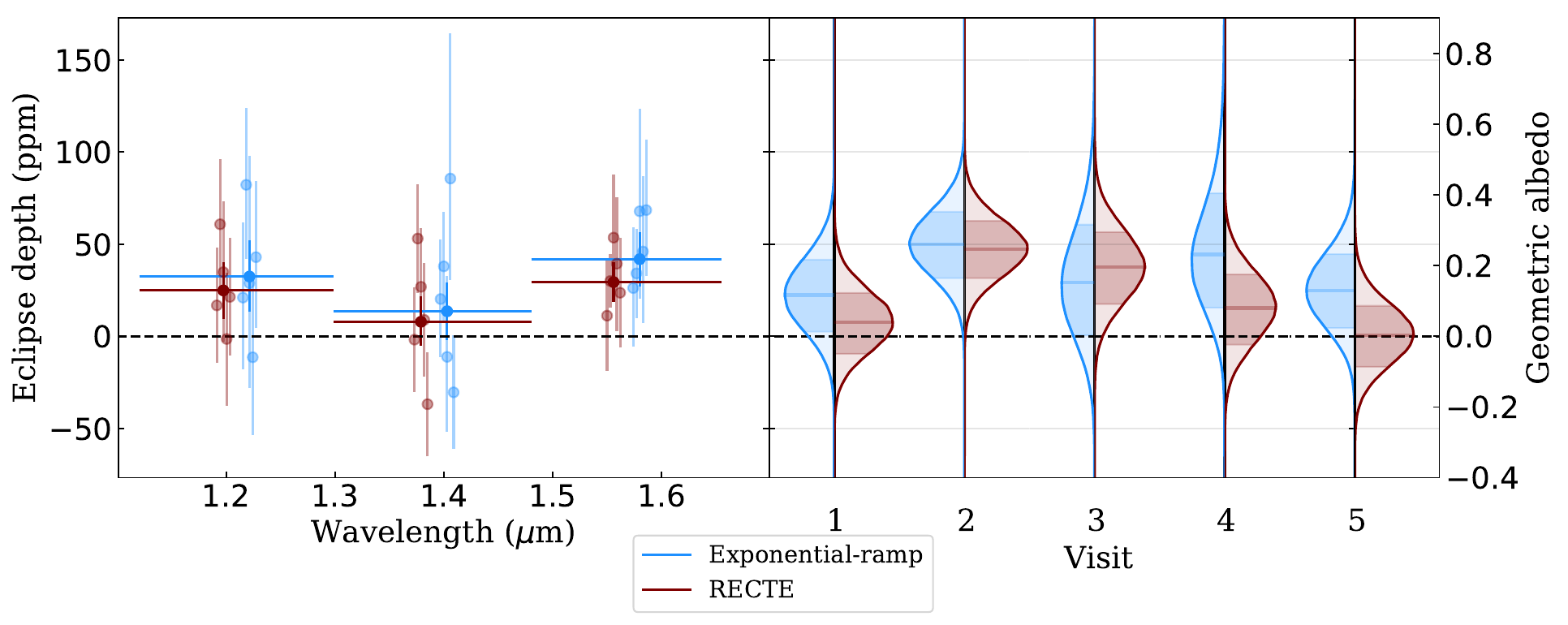}
    \caption{Near-infrared secondary eclipse depth measurements of WASP-80b for the exponential-ramp method (blue) and the \texttt{RECTE} method (maroon). The right axis shows the geometric albedo if there were zero thermal emission. \\
    \textbf{Left:} near-infrared secondary eclipse spectrum. The transparent data points in the background denote the per-visit eclipse depths and the opaque data points in the foreground are the weighted averages thereof. The data are offset slightly in wavelength for visibility purposes.\\
    \textbf{Right:} posterior distributions of the ``white light'' eclipse depth for each visit. The median and $1\sigma$ levels of the posteriors are marked with a darker shade. The two methods agree within $1\sigma$, except for the fifth visit, for which the methods differ by $1.2\sigma$. However, the eclipse depths measured with the exponential-ramp method are larger for all but one visit. The mean white light eclipse depth is $34\pm10$\,ppm for the exponential-ramp method and $22\pm9$\,ppm for the \texttt{RECTE} method.}
    \label{fig:Observations}
\end{figure*}

We used the data reduction pipeline described by \citet{Arcangeli2018} and updated by \citet{Jacobs2022} to convert the raw data into spectra. For each visit, we picked the first exposure as our reference exposure. 

\subsection{Systematics correction}
Raw WFC3 light curves show an exponential ramp in each HST orbit caused by the trapping of charges \citep{Zhou2017}. The ramps are strongest in the first orbit, which we therefore discarded. We removed the ramp in the other orbits by employing two different techniques: the empirical exponential-ramp method \citep[e.g.][]{Kreidberg2014a,Arcangeli2019,Wong2022} and the physically motivated \texttt{RECTE} method \citep{Zhou2017}. We provide a detailed description of both light-curve fitting methods in Appendix \ref{app:fitting_methods}. In both methods the system parameters are fixed to the values found by \citet{Wong2022}.

We split the WFC3 data into three spectral bins and performed the above fitting methods on all visits separately for each bin. We display a compilation of the WFC3 spectroscopic secondary eclipse light-curve fits in Appendix \ref{app:LCs}.

\subsection{Observed spectrum}
\label{sec:results_obs_spec}

The left panel of Figure \ref{fig:Observations} shows the fitted eclipse depths for each wavelength bin, method, and visit. The right panel shows the band-integrated ``white light'' eclipse depths for each visit, which are weighted averages of the spectral eclipse depths. The eclipse depths from the \texttt{RECTE} method are almost uniformly lower than for the exponential-ramp method. In Appendix \ref{app:diff_methods} we compare the \texttt{RECTE} and exponential-ramp methods and conclude that the exponential-ramp method is slightly more reliable for these data. Using the \texttt{RECTE}-reduced data does not change the results of this work significantly.

Averaging the eclipse depth over all visits yielded a total average eclipse depth in the G141 wavelength range of 
$34\pm10$\,ppm for the exponential-ramp method, implying a detection of the WASP-80b secondary eclipse at $3.4\sigma$. This makes WASP-80b the planet with the lowest equilibrium temperature for which a secondary eclipse has been detected at ${<}2.5$\,$\mu$m to date \citep{Angerhausen2015, Mansfield2021, Wong2021}. 

The geometric albedo of WASP-80b is
\begin{equation*}
A_g=(a/R_p)^2F_p/F_s=\frac{F_p/F_s}{192\rm{\,ppm}}.
\end{equation*}
Depending on the fraction of flux originating from thermal emission, we therefore measure an upper limit $A_g<0.33$ at 3$\sigma$ in the G141 wavelength range.  

\section{Atmospheric models and results}

\subsection{Atmospheric model description}
\label{sec:spec_models}
To interpret the observations, we generated one-dimensional atmospheric models that include condensate clouds and photochemical hazes. We first computed clear-atmosphere models with WASP-80b's planetary parameters using the Extrasolar Giant Planet (EGP) radiative-convective-thermochemical equilibrium code \citep{mckay1989,marley1996,fortney2005egp,Morley2012} with atmospheric metallicities of solar, $3{\times}$ solar, and $10{\times}$ solar. The resulting temperature-pressure (TP) profiles were then fed into the Community Aerosol and Radiation Model for Atmospheres \citep[CARMA;][]{turco1979,toon1988,ackerman1995,Gao2018} to act as the background atmosphere for simulations of photochemical hazes, following the setup in \citet{Gao2023}. For each metallicity, we ran soot and tholin haze models for column haze production rates between $\eta=10^{-14}$ and $\eta=10^{-9}$\,g\,cm$^{-2}$s$^{-1}$ with logarithmically spaced intervals of $10^{0.5}$. We used a constant eddy diffusion coefficient $K_{\rm{zz}}$ of $10^8$\,cm$^2$s$^{-1}$, full day-night heat redistribution and an internal temperature $T_{\rm{int}}=100$\,K. The haze optical properties were then fed back into the EGP code to calculate their feedback on the TP profiles.  Using the same background atmospheres, we used the EddySed model \citep{ackerman2001} to generate cloud distributions self-consistently, assuming clouds composed of Na$_2$S, MnS, and Cr and sedimentation efficiencies $f_{\rm sed}$ of 0.5, 1.0, and 2.0. We could not model lower sedimentation efficiencies as we were unable to make them converge given the clouds' strong impact on the TP profiles. We considered spherical haze and cloud particles and calculated their optical depth per model atmospheric layer, single-scattering albedo, and asymmetry parameter assuming Mie scattering. We subsequently used PICASO 3.0 \citep{Batalha2019,mukherjee2023} to generate reflected light, emission, and transmission spectra from the atmospheric structure and composition and aerosol optical properties.

\subsection{Atmospheric model comparison to the data}
\begin{figure*}
    \centering
    \includegraphics[width=\textwidth]{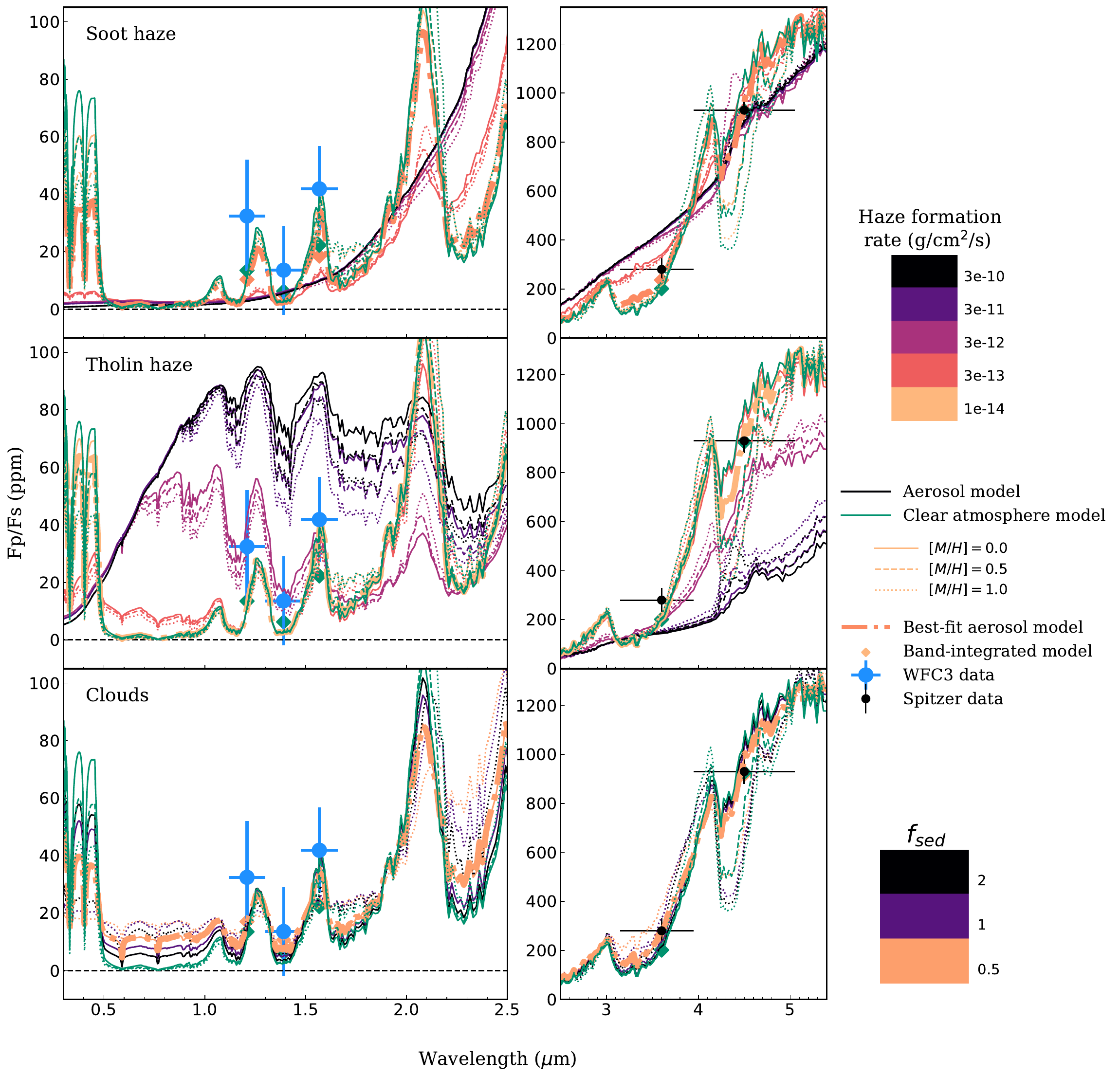}
    \caption{NIR secondary eclipse spectrum of WASP-80b measured with WFC3 (blue) and complemented with previously published Spitzer observations (black). A selection of the atmospheric models from this work are overplotted for a soot haze (upper panel), tholin haze (middle panel), and clouds (lower panel) using temperature-pressure profiles presented in Figure \ref{fig:TP-profiles}. The atmospheric models are separated into three metallicities: $[M/H]=0$ (solid), $[M/H]=0.5$ (dashed), and $[M/H]=1$ (dotted). In each panel, we also provide the clear atmosphere models in green that are mostly thermal and only contain discernible reflective features at ${<}0.6$\,$\mu$m. The best-fit models in each scenario are displayed with a bold dash-dotted line. The best-fit tholin model tracks the clear atmosphere model very closely, as the best-fit tholin production rate is low. Diamonds denote the band-integrated eclipse depths in the WFC3/Spitzer wavelength bins for the $[M/H]=0$ clear atmosphere model and for the best-fit aerosol model. It can be assumed that at ${\lesssim}2.0$\,$\mu$m any planetary flux higher than the clear atmosphere model corresponds to light reflected by aerosols.}
    \label{fig:Spectrum_and_models}
\end{figure*}

\begin{figure*}
    \centering
    \includegraphics[width=\textwidth]{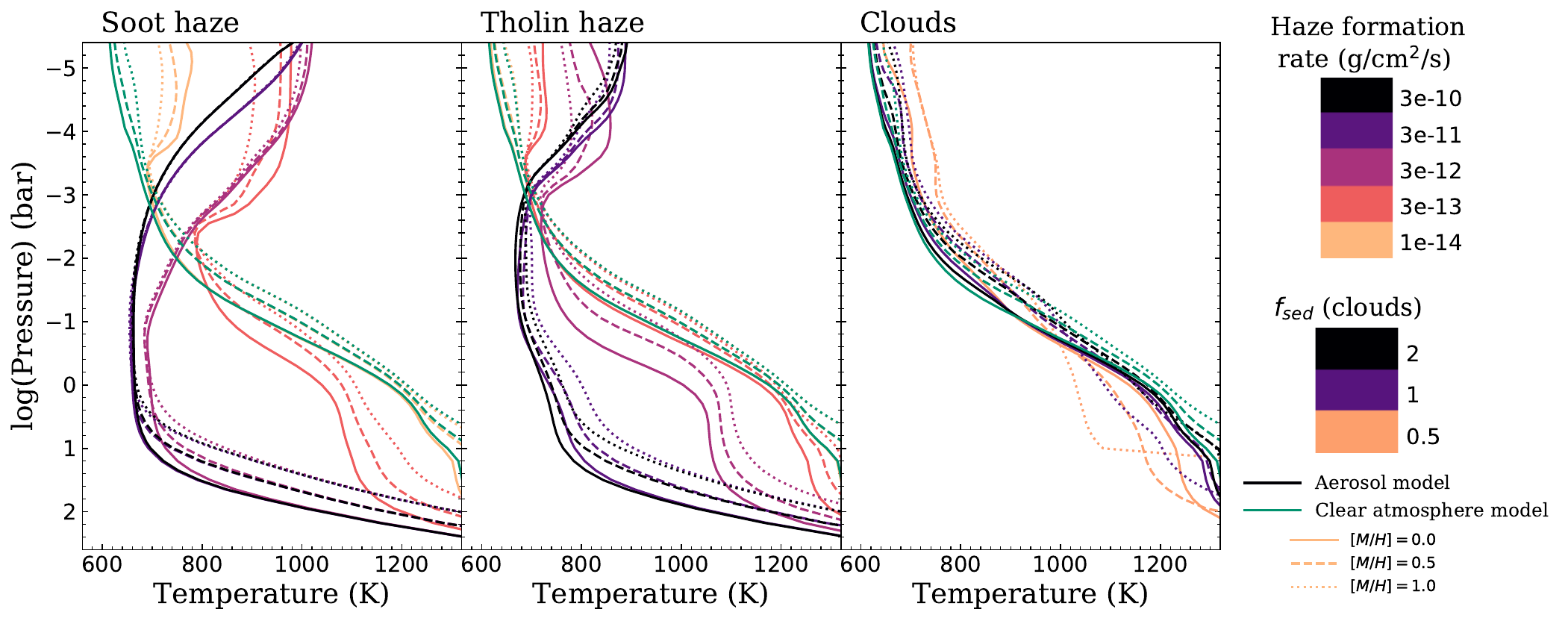}
    \caption{Temperature-pressure (TP) profiles for a selection of atmospheric models  (described in Section \ref{sec:spec_models}) for a soot haze (left), tholin haze (middle), and clouds (right). These TP profiles are used to generate the spectra in Figure \ref{fig:Spectrum_and_models}. Solid lines denote $[M/H]=0$, dashed lines show $[M/H]=0.5$ models, and dotted lines denote $[M/H]=1$. As a reference, each panel also shows in green the TP profiles of the clear atmosphere models generated with the EGP radiative-convective-thermochemical equilibrium code \citep{Morley2012}. 
    }
    \label{fig:TP-profiles}
\end{figure*}

In Figure \ref{fig:Spectrum_and_models} we complement the eclipse depths from this work with data taken by Spitzer at 3.6 and 4.5\,$\mu$m \citep{Wong2022}. We compare them to four scenarios: a clear atmosphere, an atmosphere with a soot haze, an atmosphere with a tholin haze, and a cloudy atmosphere. At wavelengths ${\lesssim}2.0$\,$\mu$m reflection by aerosols can dominate the spectrum, while these features are overwhelmed by the thermal emission at longer wavelengths. In Figure \ref{fig:TP-profiles} we display the TP profiles that produce the model spectra of Figure \ref{fig:Spectrum_and_models}.

All three WFC3 data eclipse depths are deeper than would be expected from just thermal emission from a clear atmosphere assuming $T_{\rm{int}}=100$\,K and full day-night heat redistribution (green curves in Figure \ref{fig:Spectrum_and_models}). Higher internal temperatures do not significantly alter $F_p/F_s$, but assuming zero day-night heat redistribution triples $F_p/F_s$ at WFC3 wavelengths. However, that would also double the flux at Spitzer wavelengths. The Spitzer data therefore best fit a reradiation factor \citep{Lopez2007} of 0.25 (full redistribution), with a 3$\sigma$ upper limit of 0.31. Moreover, planets near WASP-80b's $T_{\rm{eq}}$ have been observed and theorized to have near-full redistribution \citep{Komacek2017}. We therefore assume full redistribution and interpret the $1.6\sigma$ difference between model and data as due to reflection from aerosols, though we remain cautious about this model assumption.  Also visible in Figure \ref{fig:Spectrum_and_models} is how the \cotwo\ absorption feature at 4.3\,$\mu$m increases with metallicity for the clear atmosphere model. This is expected since a higher metallicity moves the balance between \cotwo/\methane production toward \cotwo\ \citep{Venot2014, Soni2023}.

The higher the haze formation rate, the more the haze model spectra deviate from the clear atmosphere model. Higher haze formation rates promote larger particle sizes and densities, leading to stronger optical absorption that creates a thermal inversion: the upper atmosphere heats up, while the deeper atmosphere cools down (see Figure \ref{fig:TP-profiles}). Greater haze formation therefore turns the \cotwo\ absorption feature at 4.3\,$\mu$m into an emission feature. The emission feature appears weak because the optically thick haze prevents us from probing deeper into the cooler atmospheric layers; as such, the spectrum resembles a blackbody at shorter wavelengths.

Stronger soot production hardly increases the planet's reflectivity because soots have a low single-scattering albedo. Conversely, an increase in tholin haze production increases the planet's reflectivity significantly. This consequently increases the flux at ${<}2.0$\,$\mu$m, but decreases the planet's flux at longer wavelengths: the heightened reflectivity reduces the total 
energy absorbed and cools down the planet, creating a spectral see-saw effect.

If there are clouds on WASP-80b's dayside, our models indicate they are likely embedded so deep into the atmosphere that their reflectivity changes the dayside spectrum only marginally. Higher sedimentation efficiencies result in a more vertically compact layer of clouds, while lower sedimentation efficiencies create a more vertically extended cloud layer, causing a slightly higher albedo.

\subsection{Model fits to the spectra}
For each scenario, we interpolated over the model grid and used a Markov Chain Monte Carlo (MCMC) sampler to find the best-fit solution and the uncertainties on the parameters. We used uniform priors within the parameter ranges described in Section \ref{sec:spec_models} and performed fits on both the dayside spectrum and the transmission spectrum from \citet{Wong2022} separately. These best-fit models are plotted in Figure \ref{fig:Best-fit-spectra} and we show the posterior distributions in Figure \ref{fig:posteriors}. We provide a more detailed accounting of the fits in Appendix \ref{app:Fit_tables}. In Figure \ref{fig:Spectrum_and_models} the best-fit models to the dayside spectrum are marked as dash-dotted lines.

In all scenarios for the dayside spectrum, a near-solar metallicity is preferred to account for the relatively small \cotwo\ absorption feature at $4.5$\,$\mu$m shown in Figure \ref{fig:Spectrum_and_models}. Although the self-consistent clear atmosphere models fit the Spitzer points well, they predict shallower eclipse depths than observed for every WFC3 wavelength bin. As such, the best-fit metallicity ($[M/H]=0.24^{+0.20}_{-0.15}$) has $\chi_{\nu}^2=1.4$. While this is a decent fit, the flux surplus at WFC3 wavelengths indicates that the NIR light detected by HST may not be purely of thermal origin, but also reflective.

The best-fit haze models tend toward low haze production (soot: log$_{10}(\eta)=-13.3\pm0.5$, tholins: log$_{10}(\eta)=-13.2^{+0.7}_{-0.6}$\,g\,cm$^{-2}$s$^{-1}$) with $3\sigma$ limits of log$_{10}(\eta)<-10.7$ for soots and log$_{10}(\eta)<-12.0$\,g\,cm$^{-2}$s$^{-1}$ for tholins. The best-fit tholin model has little to no haze formation because of the strong atmospheric cooling that reflection from tholins create.

The cloudy models have a weak preference for a low $f_{\rm{sed}}$, which increases the short-wavelength reflectivity. Cloudy models reach a $\chi^2_\nu=0.6$. Therefore, they are slightly favored over haze models by the Bayesian information criterion: $\Delta\rm{BIC}={-}2.3$ compared to a soot haze and $\Delta\rm{BIC}={-}2.9$ compared to a tholin haze. However, all aerosol models are consistent with a clear atmosphere model.

The model fits to the transmission spectrum strongly favor a high metallicity, somewhat hazy atmosphere with log$_{10}(\eta)=-12.96^{+0.12}_{-0.09}$ g\,cm$^{-2}$s$^{-1}$ for soots, and log$_{10}(\eta)=-12.13^{+0.14}_{-0.11}$ g\,cm$^{-2}$s$^{-1}$ for tholins. Hazes are required to create the optical slope. The large difference in transit depths of the two Spitzer points indicates a high \cotwo/\methane ratio and hence a high metallicity. Both the haze as well as the higher metallicity subdue the water absorption feature at $1.4$\,$\mu$m. The best-fit model is the soot haze model at $\chi^2_\nu=1.1$. A clear or purely cloudy atmosphere is excluded.

\subsection{Model comparison between dayside and limb}
\begin{figure*}
    \centering
    \includegraphics[width=\textwidth]{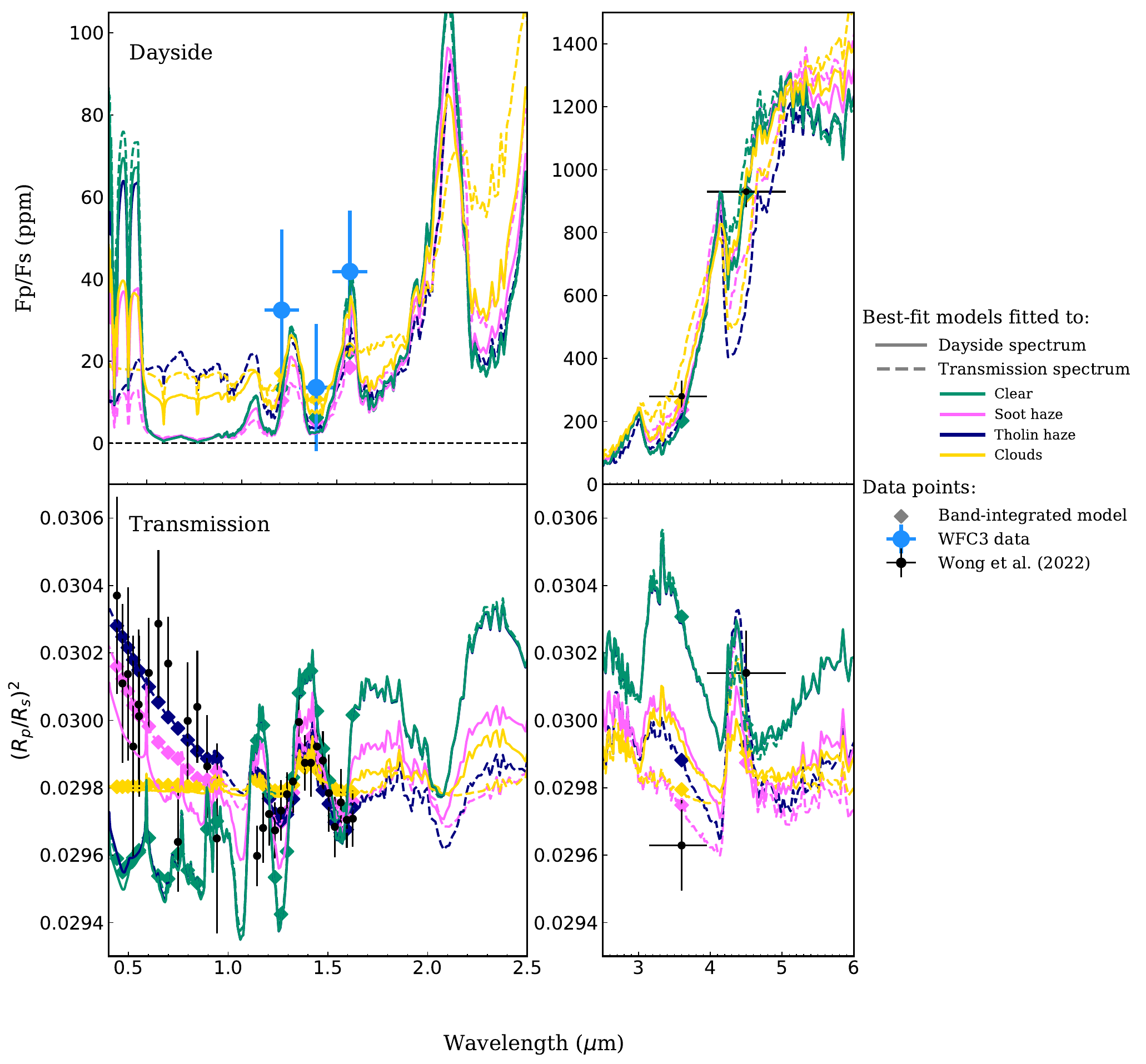}
    \caption{Best-fit models to the WASP-80b dayside (upper panel) and transmission (lower panel) spectra. Solid lines denote models fit to the dayside spectrum data. Spectra denoted with dashed lines were fit to the transmission spectrum data. All data, except for the WFC3 dayside data in blue, are adopted from \citet{Wong2022}. All data in the left panels were taken with WFC3 and the data in the right panels were taken with Spitzer. We color the clear atmosphere model green, the soot haze model pink, the tholin haze model blue, and the cloudy model yellow. Diamonds denote the band-integrated eclipse/transit depths of the model fits to the dayside spectrum. 
    }
    \label{fig:Best-fit-spectra}
\end{figure*}

\begin{figure*}
    \centering
    \includegraphics[width=\textwidth]{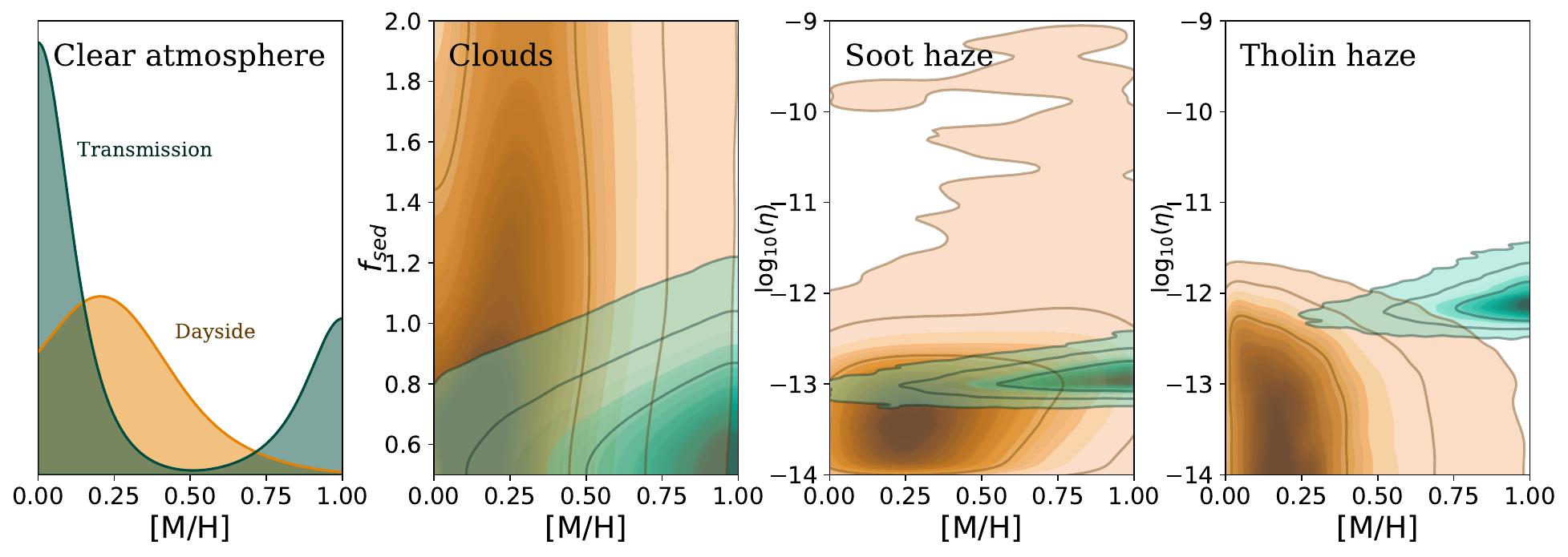}
    \caption{Posterior distributions from MCMC fits to the dayside spectrum (orange) and the transmission spectrum (green). For each model, the x-axis shows the metallicity distribution. For the cloudy model, we plot the sedimentation efficiency $f_{\rm{sed}}$ on the y-axis and for the haze models, we plot the logarithm of the haze formation rate $\eta$ (in g\,cm$^{-2}$s$^{-1}$) on the y-axis. The three isodensity lines denote the 1$\sigma$, 2$\sigma$, and 3$\sigma$-limits of the posteriors. An extra monochromatic offset parameter was used to fit transmission spectra. Correlations with that parameter are not displayed here. None of the posterior samples here have $\chi^2_{\nu}<1.6$ for both spectra simultaneously. 
    }
    \label{fig:posteriors}
\end{figure*}

We assess how well our best-fit dayside atmospheric models perform on the limb spectrum and vice versa in Figure \ref{fig:Best-fit-spectra} and Appendix \ref{app:Fit_tables}. Here we assumed the aerosol parameters to be uniform across the planet's atmosphere. Figure \ref{fig:posteriors} shows the significant overlap of the posterior distributions of all models. For all aerosol scenarios, the posteriors contain samples with $\chi^2_\nu<2.0$ for both the transmission spectrum and the dayside spectrum, indicating reasonable fits to both parts of the planet. However, none have $\chi^2_\nu<1.6$ for both spectra, which shows that no models fit both spectra well concurrently. Soot hazes generate the lowest combined $\chi^2_\nu$. Stellar activity may have steepened/induced an optical spectral slope or offsets between transmission observation epochs \citep{Rackham2018,Wong2022}. Alternatively, the true aerosol composition may be a mix between the aerosol types discussed in this work, like that found by \citet{Wong2022}, or the composition may be entirely different \citep{He2023}. The aerosol composition could also differ between the dayside and the limbs of the planet.

\section{Discussion and conclusion}
\label{sec:discussion}

Although the WFC3 dayside spectrum of WASP-80b shows signs of reflection from the planet, combining the WFC3 spectrum with the Spitzer data allows us to reject high haze formation rates $\eta<10^{-10.7}$ for soots and $\eta<10^{-12.0}$\,g\,cm$^{-2}$s$^{-1}$ for tholins. Furthermore, clouds are slightly favored over hazes.

Hazes are formed after the dissociation of, e.g. \methane, HCN, CO, \cotwo\ and C$_2$H$_2$ molecules under strong UV irradiation in the upper atmosphere (${\gtrsim}10^{-4}$\,bar) \citep{Kawashima2018,He2018a}. \methane\ has been detected on the limbs and the dayside of WASP-80b with JWST \citep{Bell2023}, and with high-resolution spectroscopy \citep{Carleo2022}; the latter report an HCN detection but remained inconclusive regarding C$_2$H$_2$. \citet{Fortney2013} show with thermochemical equilibrium models that the haze production rate is approximately highest for planets with the equilibrium temperature of WASP-80b. 
Given the host star's strong UV flux, the ingredients for strong haze formation are present at WASP-80b. Haze production on WASP-80b could be suppressed if there is a relative lack of \methane\ in the upper atmosphere due to, for example, increased eddy mixing in the deep atmosphere  \citep{Baxter2021}.

In addition to the UV flux, haze production may increase with increasing temperatures \citep{He2020}. The dayside receives both higher UV fluxes and hosts higher temperatures than the limbs. Yet, the inferred dayside haze production is similar to, or possibly even lower than, the limb haze production, which we probe with the transmission spectrum. Hazes produced on the dayside may be rapidly advected to the evening terminator by an eastward jet. The evening limb could therefore show a higher haze mass than the dayside \citep{Steinrueck2021}.

\citet{Wong2022} infer a ${\sim}30$--$100$ times solar metallicity from low-resolution transit spectroscopy because of a weak water feature and a large $4.5$\,$\mu$m transit depth that may be indicative of strong \cotwo\ absorption.  Conversely, \citet{Carleo2022} suggest an atmosphere that is consistent with a solar composition from high-resolution transit spectroscopy. In line with \citet{Wong2022}, our transmission spectrum fits tend toward high metallicities. However, our dayside spectrum fits tend toward solar metallicities because the relative flux in the Spitzer bandpasses does not indicate a strong \cotwo\ feature.
As \citet{Wong2022} note, a possibly 
anomalous 4.5\,$\mu$m Spitzer transmission data point may cause this difference. Excluding this point would allow for solar compositions with both the models from \citet{Wong2022} and our models. The best-fit aerosol composition is not significantly different at solar metallicities. 

\citet{Wong2022} analyzed the WASP-80b transmission spectrum with a free-retrieval. They inferred a fine-particle (${\lesssim}0.1$\,$\mu$m) haze with a deep cloud deck at $\chi^2_\nu=0.88$. Our grid-retrieval over atmospheric models was able to find almost equally good fits ($\chi^2_\nu=1.1$) with only a haze. The average particle size in our CARMA models matches that inferred by \citet{Wong2022}.

Although the modeled sedimentation efficiency of clouds has a marginal impact on the dayside spectrum, the results show a slight preference for low sedimentation efficiencies ($f_{\rm{sed}}<1$), in line with previous works \citep{Morley2013,Adams2022}. Those works find even lower values for $f_{\rm{sed}}$ than we could self-consistently model here. Theoretically, lower values of $f_{\rm{sed}}$ would increase the albedo, potentially improving the fit to the WFC3 data.

Comparing the retrieved aerosol composition from WASP-80b's dayside and transmission spectra shows a possible difference in aerosol composition and/or abundance between the dayside and limbs of the planet. Alternatively, the data can be interpreted with a uniform soot haze model, which fits both spectra at $\chi_\nu^2=1.6$. A comprehensive joint analysis of both the transmission and the secondary eclipse spectra can significantly help disentangle degenerate solutions to the data 
\citep{Griffith2014, Zhang2020}. However, such analysis is outside the scope of this paper due to the one-dimensional nature of the atmospheric models and the limited precision and spectroscopic abilities of WFC3 and Spitzer. The GTO observations that are being taken by JWST for WASP-80b are well suited to such combined analysis. 

\begin{acknowledgements}
J.M.D acknowledges support from the Amsterdam Academic Alliance (AAA) Program, and the European Research Council (ERC) European Union’s Horizon 2020 research and innovation program (grant agreement No. 679633; Exo-Atmos). This work is part of the research program VIDI New Frontiers in Exoplanetary Climatology with project number 614.001.601, which is (partly) financed by the Dutch Research Council (NWO). Support for program HST-GO-15131 was provided by NASA through a grant from the Space Telescope Science Institute, which is operated by the Associations of Universities for Research in Astronomy, Incorporated, under NASA contract NAS5-26555.
\end{acknowledgements}

%
%

\bibliography{references}{}
\bibliographystyle{aasjournal}
\appendix
\restartappendixnumbering

\section{Light-curve fitting methods}
\label{app:fitting_methods}
To remove the telescope systematics, we employ two different methods: the exponential-ramp method \citep[e.g.][]{Kreidberg2014a, Arcangeli2019}, and the \texttt{RECTE} method \citep{Zhou2017}. An exponential-ramp method was also used for the transmission spectroscopy measurements of WASP-80b by \citet{Wong2022}.

\subsection{Exponential-ramp method}
\label{app:fitting_methods_exp}
In the exponential-ramp method one fits an analytical model, to the light curve of each visit. The model has the form:
\begin{eqnarray}
\label{eq:xi}
F(\mathbf{t}) &=& M_{\lambda, v}(\mathbf{t})\, \Theta_{\lambda, v, s}(\mathbf{t}) \,T_{\lambda, v, o}(\mathbf{t}) \\
\label{eq:xi2}
\Theta_{\lambda, v, s}(\mathbf{t}) &=& C_{\lambda, v, s} + V_{\lambda, v, s}(\mathbf{t} - t_{\rm{ecl}})\\
T_{\lambda, v, o}(\mathbf{t}) &=& 1 - R_{\lambda, v, o} e^{-(\mathbf{t} - t_b) / \tau_{\lambda, v}}
\end{eqnarray}
where $M(\mathbf{t})$ is the secondary eclipse model from \texttt{batman} \citep{Kreidberg2015}, $\mathbf{t}$ is a vector of observation times, $C$ is a normalization constant, $V$ is a visit-long linear slope, $R$ is the WFC3 ramp amplitude, $\tau$ is the ramp timescale, $t_b$ is the time of the first exposure in the orbit, and $t_{\rm{ecl}}$ is the mid-eclipse time. The subscripts $\lambda, v, s$ and $o$ denote whether a parameter is a function of wavelength, telescope visit, scan direction, and orbit number respectively. In this work, the telescope systematics were not shared between visits because the observational orientation and the number of subexposures per exposure varies per visit (see Table \ref{tab:obs}). We set the ramp amplitude of the last orbit equal to the amplitude of the penultimate orbit.

\begin{deluxetable}{llll}[h!]
\tablewidth{0pc}
\setlength{\tabcolsep}{3pt}
\tabletypesize{\small}
\tablecaption{ Observation Log \label{tab:sample}}

\tablehead{
Visit & UT start date  & $n_{\rm{sub}}$ \tablenotemark{\scriptsize a} & $\phi$ \tablenotemark{\scriptsize b} } 
\startdata
1     & 17 June 2019   & 13                 & 40                                                                                                                  \\
2     & 5 July 2019    & 12                 & 40                                                                                                                  \\
3     & 11 July 2019   & 12                 & 31                                                                                                                  \\
4     & 8 August 2019  & 12                 & 330                                                                                                                 \\
5     & 11 August 2019 & 13                 & 330                     
\enddata
\textbf{Notes.}
\vspace{-0.2cm}\tablenotetext{\textrm{a}}{ Number of subexposures per exposure.}
\vspace{-0.2cm}\tablenotetext{\textrm{b}}{ Position angle of V3-axis of \textit{HST} (deg).}
\vspace{-0.2cm}

\label{tab:obs}
\end{deluxetable}
There is no variation of the measured eclipse depths with the position angle of the telescope. However, eclipses measured in the 13-subexposure mode are $1.5\sigma$ shallower than the eclipses measured in the 12-subexposure mode. 
We removed the 13-th subexposure from the first and fifth visits and reran the fits. This did not change the eclipse depths significantly (${<}5$\,ppm). 
The variations in eclipse depth may, therefore, originate from the variability of the host star.

The exponential-ramp method is often unable to correctly fit for the first exposure of each orbit \citep[e.g.][]{Changeat2021, Gressier2022} as it often deviates from the ramp model. We therefore removed them for this method.

\subsection{\texttt{RECTE} method}
In the \texttt{RECTE} method, we swap $T(t)$ in Equation \ref{eq:xi} for the system of differential equations from \citet{Zhou2017} that govern the filling and emptying of the charge traps. This requires the fitting of four other parameters: the number of initially filled fast/slow charge traps ($E_{f/s,0_{v,\lambda}}$) and the number of fast/slow charge traps that are filled in between orbits ($\Delta E_{f/s_{v,\lambda}}$). These parameters are the same for each scan direction, but different for each visit and wavelength. Contrary to the exponential-ramp model, the \texttt{RECTE} method is able to cope with the low flux values of the first exposure of each orbit. For this method, we therefore leave these exposures in our fits.

The \texttt{RECTE} modeling approach may enable the use of the first orbit of every visit that would otherwise be discarded \citep{Zhou2017}. We experimented with maintaining this extra orbit in our analysis as that could decrease the measurement uncertainty. However, we found this particular data set to be ill-suited to such analysis with \texttt{RECTE}. The orbit-long ramp of the first orbit is so strong that any good fit with \texttt{RECTE} requires negative initially filled charge traps, which are unphysical. Limiting $E_{f/s,0_{v,\lambda}} \geq 0$ yields bad fits (on average 115\% above photon noise) and strongly negative eclipse depths at a mean white light eclipse depth of $-125\pm7$\,ppm. We therefore decided to remove the first orbit of each visit from our \texttt{RECTE} analysis and work with the three remaining orbits, analogous to the exponential-ramp method.

\subsection{Estimation of errors}
In order to estimate the errors on our fitted parameters and identify the degeneracies in the model, we used an MCMC approach using the open-source \texttt{emcee} code \citep{emcee}. 
Each chain has 25,000 steps with 70 walkers and a burn-in period of 2000 steps.
Our average final precision on the spectroscopic eclipse depths per visit with the exponential-ramp and \texttt{RECTE} methods were 41\,ppm and 31\,ppm, respectively, per wavelength bin. We reached an average spectrophotometric precision on the best fits of 4\% and 6\% above photon noise, respectively.

\subsection{Spectral drift on the detector}
Some WFC3 light-curve fits may benefit from adding an additional term $(1 + c \ \delta_\lambda)$ to equation \ref{eq:xi2} \citep[e.g.][]{Haynes2015, Wakeford2016}, where $c$ is an additional fit parameter and $\delta_\lambda$ is the band-integrated drift of the spectrum on the detector in the dispersion direction. We measured $\delta_\lambda$ by cross-correlating the observed spectral flux of each exposure to the reference exposure. However, including fit parameter $c$ did not improve the Bayesian information criterion. We therefore did not use this term.

\section{Light curves}
\label{app:LCs}
\begin{figure*}
    \centering
    \includegraphics[width=\textwidth]{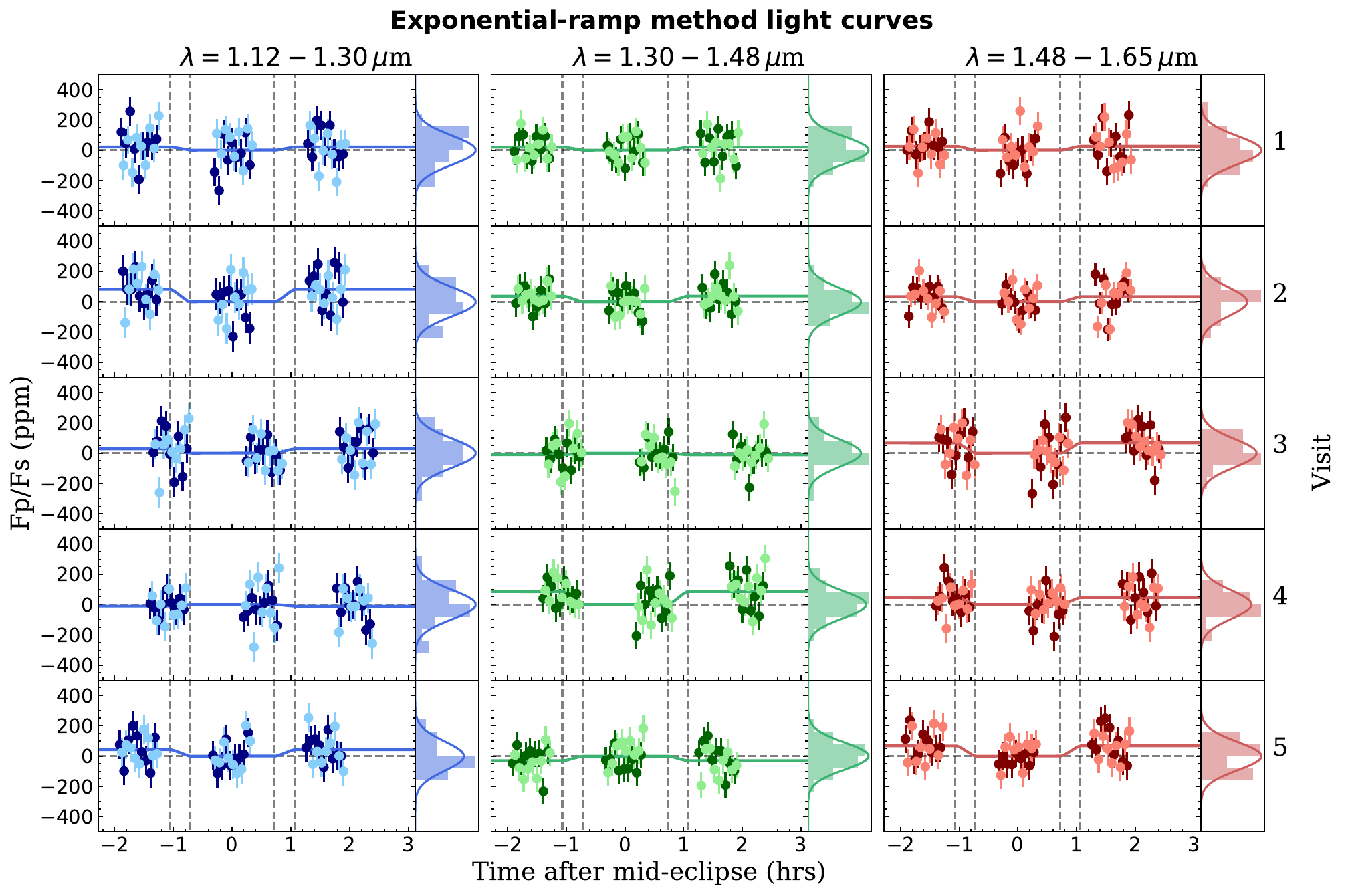}
    \caption{Measured light curves of the WASP-80b secondary eclipse divided into three wavelength bins (columns) over five visits (rows) for the exponential-ramp method. In each cell, we show the normalized, telescope systematics-reduced data and split them into forward-scanned (dark), and reverse-scanned (light) data. We determined a unique eclipse depth for each bin and each visit. The fitted light curve is shown as a solid line and we denote the times of first, second, third, and fourth contact with vertical dashed lines. For each fit, we provide a residual histogram to the right of the light curve. We compare them to the expected photon noise (solid line). The residuals satisfy a Shapiro-Wilks test at the $\alpha=0.1$ level and they can thus be considered gaussian. 
    In Figure \ref{fig:LCs_recte} we display the equivalent of this figure for the \texttt{RECTE} method. Note that for the exponential-ramp method, the first exposure of each orbit has been removed, whereas this exposure is kept for the \texttt{RECTE} method.}

    \label{fig:LCs_exp}
\end{figure*}

\begin{figure*}
    \centering
    \includegraphics[width=\textwidth]{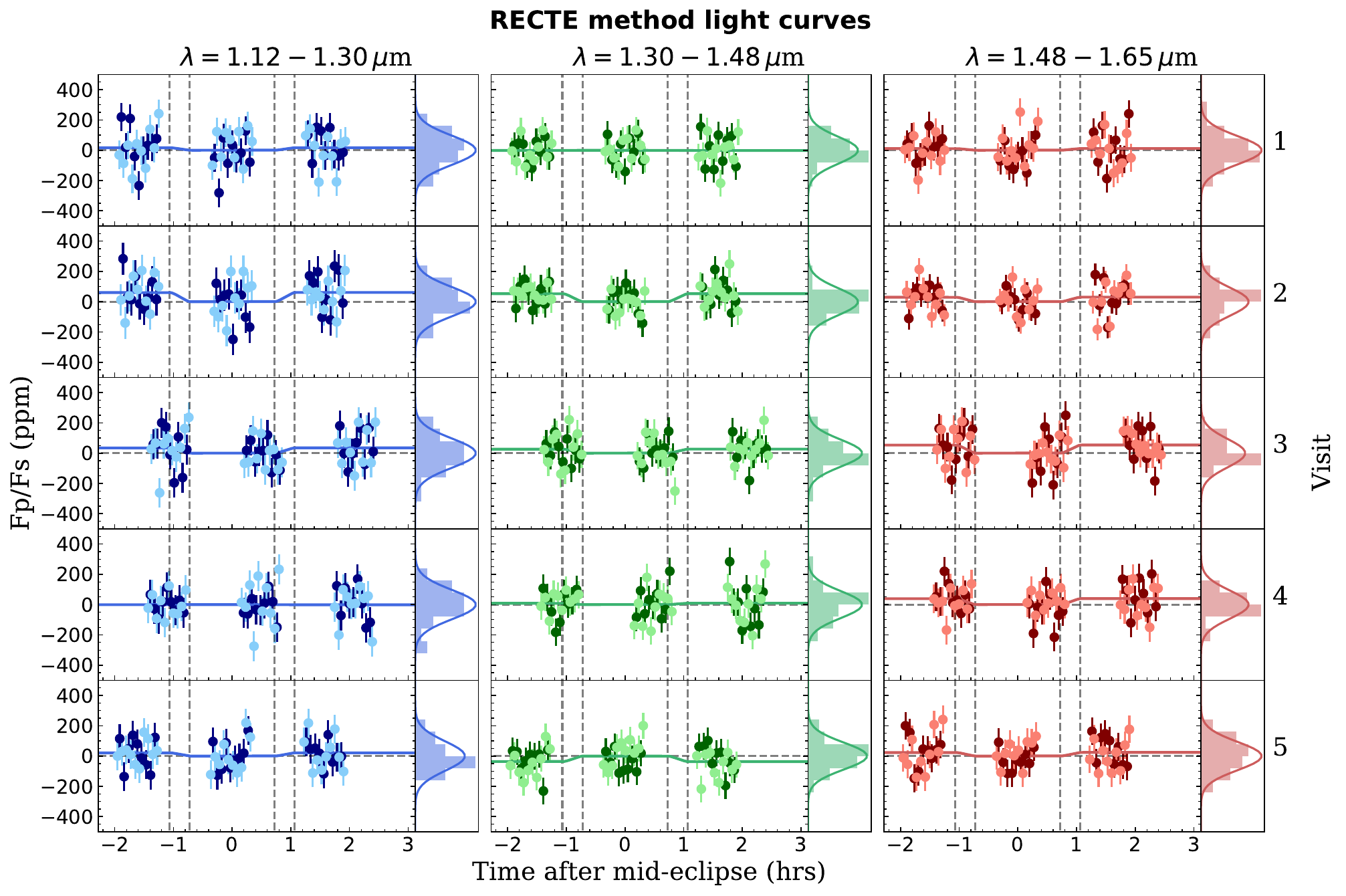}
    \caption{The same as Figure \ref{fig:LCs_exp} but for the \texttt{RECTE} method. The residuals satisfy a Shapiro-Wilks test at the $\alpha=0.15$ level and they can thus be considered gaussian. }
    \label{fig:LCs_recte}
\end{figure*}

Figures \ref{fig:LCs_exp} and \ref{fig:LCs_recte} show the secondary eclipse light curves for the exponential-ramp method and the \texttt{RECTE} method, respectively. We fit the light curves in the three wavelength bins and for each visit separately. 

We note that the scatter around the mean of the per-visit eclipse depths is 25\% smaller than the average uncertainty. For a sample size of 15 normally distributed measurements, there is a 13\% probability that their standard deviation is ${>}25$\% smaller than their uncertainty. This could be an indication that the uncertainties are overestimated. 

\section{Comparing fitting methods}
\restartappendixnumbering
\label{app:diff_methods}
It is noticeable from Figure \ref{fig:Observations} that the \texttt{RECTE} method measures shallower eclipses than the  exponential-ramp method across all three wavelength bins and for most visits. The averaged white light eclipse depth is slightly more than $1\sigma$ larger for the exponential-ramp method. In this section, we dive deeper into the performance of both methods.


Contrary to the \texttt{RECTE} method, the exponential-ramp method (see Appendix \ref{app:fitting_methods}) has one parameter, $R_{\lambda, v, 1}$, that acts only on the first orbit, hence on fewer data points than all the \texttt{RECTE} parameters act on. The uncertainty on $R_{\lambda, v, 1}$ is therefore relatively large and some degeneracies arise between it and the eclipse depth parameter. The exponential-ramp method therefore has a ${\sim}10$\% lower precision on the eclipse depth. This effect is exacerbated for visits 3 and 4. Those visits have a shorter preeclipse baseline (see Figures \ref{fig:LCs_exp} and \ref{fig:LCs_recte}) and therefore have even larger degeneracies between eclipse depths and $R_{\lambda, v, 1}$. This results in ${\sim}30$\% larger uncertainties for these two visits than for the other visits (see Figure \ref{fig:Observations}). However, the \texttt{RECTE} method having a higher precision does not necessarily mean it has a higher accuracy.

\begin{figure*}
    \centering
    \includegraphics[width=\textwidth]{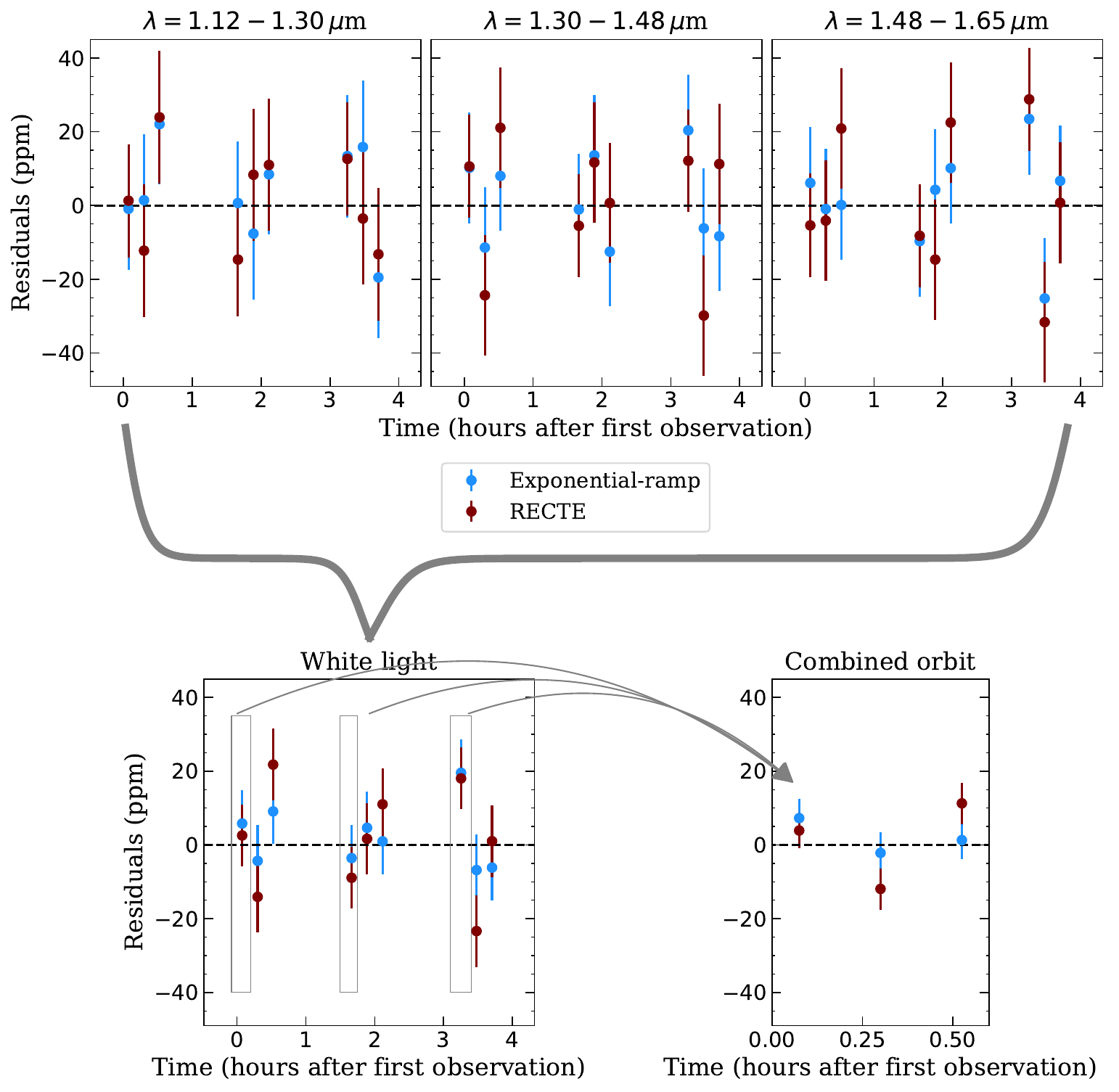}
    \caption{An investigation into the residual structures of the exponential-ramp method (blue) and the \texttt{RECTE} method (maroon).\\
    \textbf{Upper panels:} residuals of the exponential-ramp light curves (from Figure \ref{fig:LCs_exp}) and the \texttt{RECTE} light curves (from Figure \ref{fig:LCs_recte}) averaged over all visits. We combined the first seven exposures, middle six exposures, and the last six exposures of each orbit into three separate data points per orbit.\\ 
    \textbf{Lower left:} residuals of the upper panels averaged over wavelength.  \\
    \textbf{Lower right:} averaged residual structure for an orbit. That is, to obtain the first data point we averaged over the first data point, of each orbit in the lower left panel; to obtain the second data point we averaged over the second data point of each orbit the lower left panel; and for the third, we averaged over the third data point of each orbit. The lower level of scatter in this diagram for the exponential-ramp method shows that that method is more capable of capturing the shape of the orbit-long telescope systematics for this particular set of observations. }
    \label{fig:residual_structure}
\end{figure*}

In Figure \ref{fig:residual_structure} we explored the residual structure for both light-curve fitting methods (exponential-ramp, blue; \texttt{RECTE}, maroon). We did this by averaging the first seven exposures of the first orbit over all visits into a single data point, averaging the middle six exposures, and averaging the last six exposures into a single data point. We performed this averaging for each orbit and each wavelength bin to obtain the upper panel of Figure \ref{fig:residual_structure}. The standard deviation of all the residuals in the upper panel of Figure \ref{fig:residual_structure} is 12\,ppm for the exponential-ramp method and 16\,ppm for the \texttt{RECTE} method. This signifies that the \texttt{RECTE} method creates a larger residual structure. This is corroborated by the lower right panel, where we averaged over wavelength and over the three orbits to obtain a residual blueprint for an orbit with both methods. This is the average residual structure for an orbit. Once again, we see that the \texttt{RECTE} method creates a significantly larger intraorbit residual structure than the exponential-ramp method. The standard deviation of the residuals in the lower right panel of Figure \ref{fig:residual_structure} is 10\,ppm for the \texttt{RECTE} method and 4\,ppm for the exponential-ramp method.

Because of the above arguments and the fact that the exponential-ramp method's light curves are also closer to photon noise (4\% above it versus 6\% above photon noise), we opt to work with the exponential-ramp method as our main result.

\section{Details of the model fits}
\label{app:Fit_tables}
Table \ref{tab:chi2s} shows a detailed accounting of the model fits. These model fits are the results of fits to the dayside spectrum as well as fits to the transmission spectrum. We computed the $\chi^2_{\nu}$ on both the dayside spectrum as well as the transmission spectrum for both types of fits.

\begin{deluxetable*}{l||ll|l|l|l|l}
\tablewidth{0pc}
\setlength{\tabcolsep}{3pt}
\tabletypesize{\small}
\tablecaption{Summary of model fits. \label{tab:chi2s} }

\tablehead{
Model & \multicolumn{2}{c|}{$\chi^2_\nu$} & $\Delta$BIC \tablenotemark{\scriptsize a} & Best-fit    & Posterior Median  & $3\sigma$ Limits \\
\cline{2-3}
 & Dayside & Transmission & &Parameters\tablenotemark{\scriptsize b} & Values\tablenotemark{\scriptsize c} & \\
 & Spectrum & Spectrum  & & & &}
\startdata
 & \multicolumn{6}{c}{Models Fit to the Dayside Spectrum}\\
\cline{1-7}
Clear & $1.4$ & $6.5$\tablenotemark{\scriptsize d} & $0$ & $[M/H]=0.20$ & $[M/H]=0.24^{+0.20}_{-0.15}$ & $[M/H]<0.91$  \\
Clouds & $0.6$ & $2.0$ & $-1.3$ & $[M/H]=0.03$ & $[M/H]=0.30^{+0.24}_{-0.20}$ & $[M/H]<0.94$  \\
 & & & & $f_{\rm{sed}}=0.50$ & $f_{\rm{sed}} =1.19^{+0.6}_{-0.5}$ & $f_{\rm{sed}}$ is unconstrained  \\
Soot haze & $1.2$ & $2.3$  & $+1.0$ & $[M/H]=0.21$ & $[M/H]=0.33^{+0.4}_{-0.22}$ & $[M/H]<0.99$  \\
& &  & & log$_{10}(\eta)=-13.5$ &  log$_{10}(\eta)=-13.3\pm0.5$ & log$_{10}(\eta)<-10.7$ \\
Tholin haze & $1.3$ & $6.5$ &  $+1.6$ & $[M/H]=0.20$ & $[M/H]=0.22^{+0.20}_{-0.15}$ & $[M/H]<0.89$  \\
 &  & &   & log$_{10}(\eta)$$=-14.0$ & log$_{10}(\eta)=-13.2^{+0.7}_{-0.6}$ & log$_{10}(\eta)<-12.0$  \\
 \cline{1-7}
 & \multicolumn{6}{c}{Models Fit to the Transmission Spectrum}\\
 \cline{1-7}
Clear & $1.6$\tablenotemark{\scriptsize e} & $6.2$ & $0$ & $[M/H]=0.0$ & $[M/H]=0.12^{+0.8}_{-0.10}$ & $[M/H]$ is unconstrained \tablenotemark{\scriptsize f}   \\
Clouds & $4.2$ & $1.8$ & $-136$ & $[M/H]=1.0$ & $[M/H]=0.84^{+0.12}_{-0.25}$ & $[M/H]>0.06$  \\
 & & & & $f_{\rm{sed}}=0.50$ & $f_{\rm{sed}} =0.57^{+0.10}_{-0.06}$ & $f_{\rm{sed}}<1.0$  \\
Soot haze & $2.9$ & $1.1$ & $-156$ & $[M/H]=1.0$ & $[M/H]=0.84^{+0.12}_{-0.22}$ & $[M/H]>0.12$  \\
& &  & & log$_{10}(\eta)=-12.97$ &  log$_{10}(\eta)=-12.96^{+0.12}_{-0.09}$ & $-13.22<$log$_{10}(\eta)<-12.59$  \\
Tholin haze & $7.4$ & $1.1$ &  $-156$ & $[M/H]=1.0$ & $[M/H]=0.91^{+0.07}_{-0.14}$ & $[M/H]>0.40$  \\
 &  & &   & log$_{10}(\eta)=-12.13$ & log$_{10}(\eta)=-12.13^{+0.14}_{-0.11}$ & $-12.41<$log$_{10}(\eta)<-11.65$  
\enddata
\textbf{Notes.}
\vspace{-0.2cm}\tablenotetext{\textrm{a}}{ The difference in Bayesian information criterion with respect to the clear atmosphere model. This is performed on the spectrum to which the model was fit. A lower BIC indicates a better fit. A $\Delta\rm{BIC}{<}2$ can be deemed insignificant.}
\vspace{-0.2cm}\tablenotetext{\textrm{b}}{ Parameters used for the best-fit spectra in Figure \ref{fig:Best-fit-spectra}. The haze formation rate $\eta$ is given in g\,cm$^{-2}$s$^{-1}$. }
\vspace{-0.2cm}\tablenotetext{\textrm{c}}{ Median values of the MCMC posterior distributions. They can deviate significantly from the best-fit values because the model grid parameter space is limited and the posteriors are therefore truncated and asymmetrical. The model parameter space is limited to ($0\leq[M/H]\leq1$), ($0.5 \leq f_{\rm{sed}} \leq 2$), and ($-14 \leq \rm{log}_{10}(\eta) \leq -9$).}
\vspace{-0.2cm}\tablenotetext{\textrm{d}}{ We generated a transmission spectrum from the atmospheric model that best fits the dayside spectrum, and compared it to the HST/Spitzer transmission data.}
\vspace{-0.2cm}\tablenotetext{\textrm{e}}{ We generated a dayside spectrum from the atmospheric model that best fits the transmission spectrum, and compared it to the HST/Spitzer dayside data.}
\vspace{-0.2cm}\tablenotetext{\textrm{f}}{ The metallicity posterior of the clear atmosphere model is double-peaked at $[M/H]=0$ and $[M/H]>1$ (see Figure \ref{fig:posteriors}). A solar metallicity composition fits the water spectrum in the G141 wavelength range well, while higher metallicities suit the \cotwo/\methane ratio probed by the longer wavelength Spitzer points better.}

\end{deluxetable*}

\end{document}